\input{aipcheck}

\documentclass[
    ,final            
  ]
  {aipproc}

\layoutstyle{6x9}

\begin{document}

\title{Directional Anisotropy of \textit{Swift} Gamma-Ray Bursts}


\author{P\'eter Veres}{
  address={E\"otv\"os University, Budapest},
  altaddress={Bolyai Military University, Budapest}
}

\author{Zsolt Bagoly}{
  address={E\"otv\"os University, Budapest}
}
\author{Istv\'an Horv\'ath}{
  address={Bolyai Military University, Budapest}
}

\author{Lajos G. Bal\'azs}{
  address={Konkoly Observatory, Budapest}
}
\author{Attila M\'esz\'aros}{
  address={Charles University, Prague}
}

\author{J\'anos Kelemen}{
  address={Konkoly Observatory, Budapest}
}
\begin{abstract}
 Swift satellite measurements contributed substantially to the gamma-ray burst
 (GRB) redshift observations through fast slewing to the source of the GRBs.
 Still, a large number of bursts are without redshift.  We study the celestial
 distribution of bursts with various methods and compare them to a random
 catalog using Monte-Carlo simulations. We find an anisotropy in the
 distribution of the intermediate class of bursts and find that the short and
 long population are distributed isotropically.  
\end{abstract}
\maketitle
\section{Introduction}
		  Swift-BAT can observe $1.4$~sr of the sky at any given moment. Its
field of view is not uniform in sensitivity.  We search for directional
anisotropies of the GRBs using coordinate-system independent tests.  To compute
the significance of the tests we perform Monte-Carlo simulations for the short,
intermediate and long populations of the bursts taking into account  the
exposure function of \textit{Swift}. The group memberships are calculated on
the hardness-duration plane using a modified maximum likelihood method \citep{2010arXiv1010.2087V}.   While the exposure map of \textit{BATSE}
\citep{1992AIPC..265..399B} depends only on the declination, \textit{Swift} has
a more complicated exposure function.  To deal with this, we used the HEALPix
\citep{2005ApJ...622..759G} pixelization algorithm.
\section{Exposure function}
\begin{figure}[htb]
          \centering
          \includegraphics[width=.3\columnwidth,angle=90]{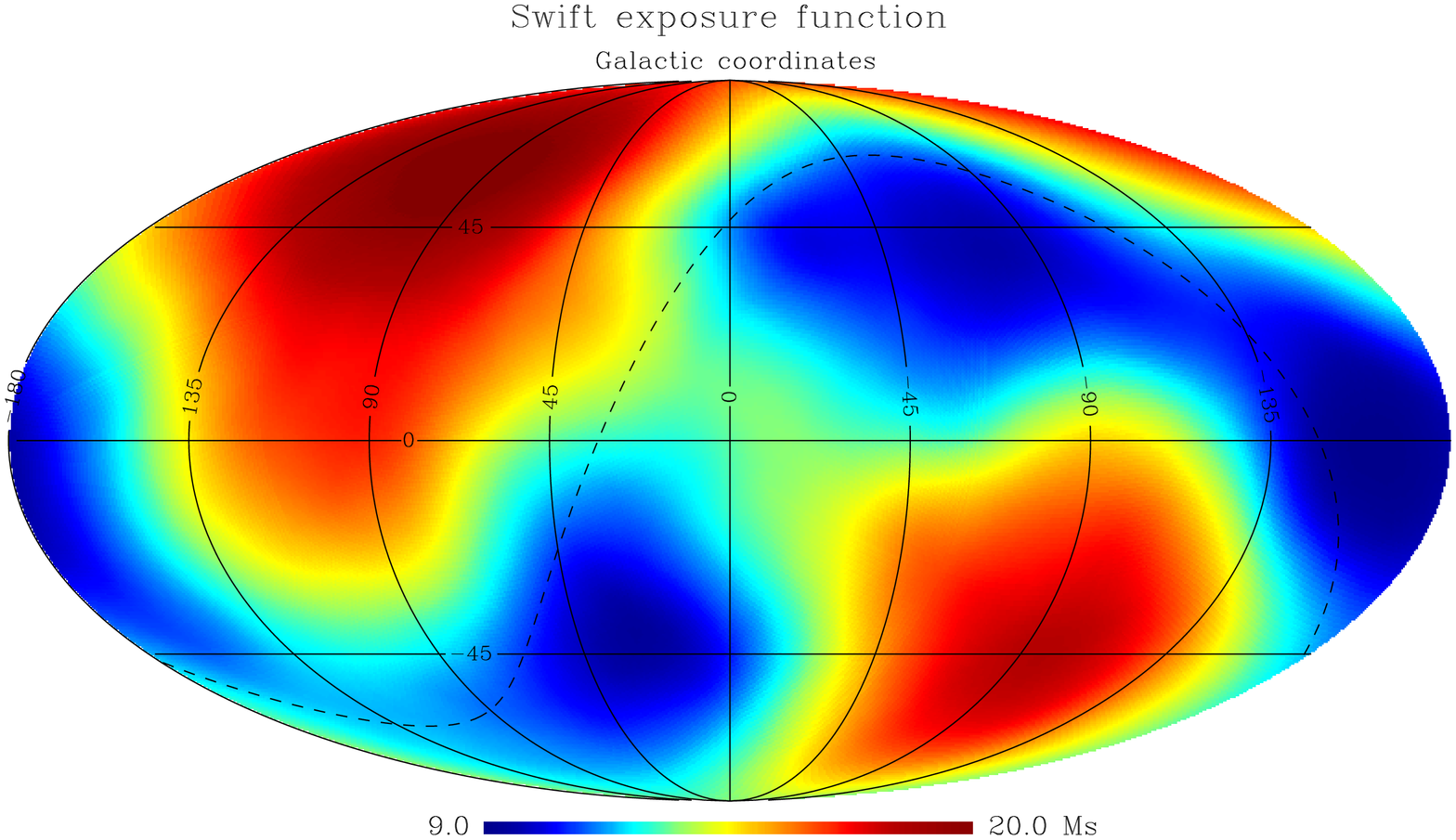}
          \includegraphics[width=.5\columnwidth]{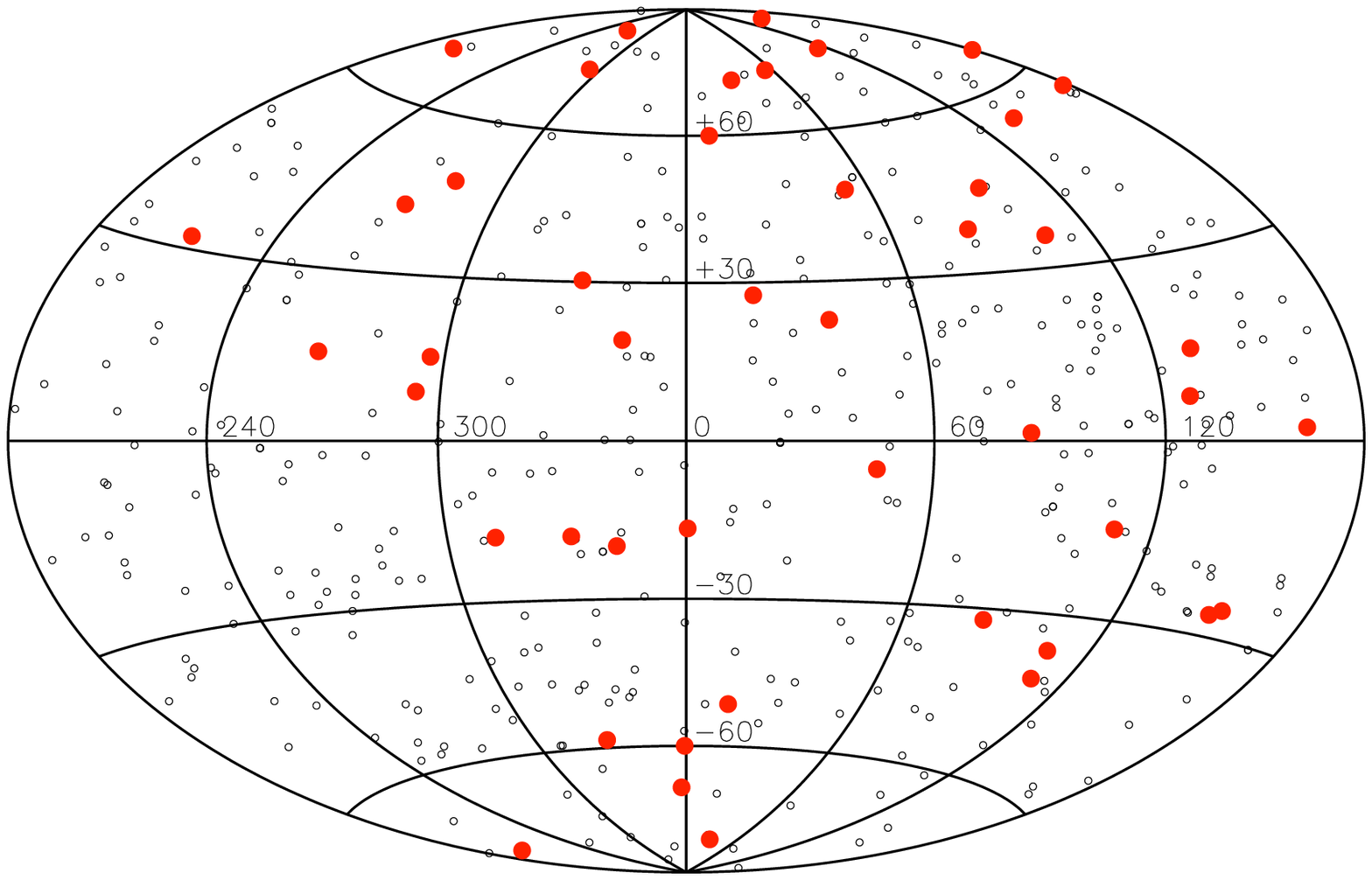}
		  \caption{Figures showing the Swift exposure function in Mollweide
		  projection in Galactic coordinates(left). The dotted line shows the
		  ecliptic plane.
		  The celestial distribution of the intermediate group (filled red
		  circles) and the total sample (empty black circles)
		  (right).}\label{exp}
           \end{figure}

	 Because Swift makes pointed observations, its sky sensitivity map will be
	 dependent on many factors.  We place the Swift sensitivity mask on the
	 celestial sphere centered on the {\it ($\alpha, \delta$)} coordinates in
	 the catalog. We rotate the mask with the appropriate angle as indicated by
	 the {\it Roll\_angle} parameter. Then we multiply this with the exposure
	 time spent on that location. We carry out this exercise for all
	 observations and sum up the results. The exposure map on Fig. \ref{exp}
	 shows a paucity of exposure time in the direction of the ecliptic, and a
	 variation of roughly a factor of two between the extremes. It also shows a
	 similarity in structure  with the 22 month exposure map of
	 \citet{2010ApJS..186..378T}.
\section{Tests of isotropy}
        Here we use the tests put forward by \citet{1993ApJ...407..126B}. The test statistics are the following:
        \[ M_N =\frac{1}{N} \sum_{i=1}^N\left[ \begin{array}{ccc}
            x_i x_i &x_i y_i  & x_i z_i \\
            y_i x_i &y_i y_i  & y_i z_i \\
			z_i x_i &z_i y_i  & z_i z_i  \end{array} \right] \hspace{1cm} R=  \sum_{i=1}^N {\bf r_i} \hspace{1cm} B= \frac{15N}{2} \sum_{k=1}^3 \left( \lambda_k-\frac{1}{3} \right)^2\]
     Here $N$ is the number of elements, $x_i, y_i$ and $z_i$ are the Cartesian
     coordinates of the bursts on the unit sphere, {\bf $r_i$} is the unit
     vector pointing to the bursts and $\lambda_k$ are the eigenvalues of
	 $M_N$ ($\lambda_1 \ge \lambda_2 \ge \lambda_3$). These statistics are
	independent of the coordinate system. $R$ is the Rayleigh and $B$ is the 
    Bingham statistic. $W$ is the Watson statistic, related simply to $R$ as $M=3R^2/N$.
\vspace{-0.5cm}
\section{Results}
We calculated the test statistic values for the actual observed distribution of
the Swift bursts. Afterwards we simulate $1000$  catalogues according to the
exposure function for the different populations and calculate the mentioned
statistics.  We find the short and long population are distributed
isotropically (p-values from $0.29$ to $0.694$ for the hypothesis of isotropic
distribution). The intermediate population however, shows a marked anisotropy
(p-values from $0.038$ to $0.072$). (see Table 1.) In the distribution of the
intermediate duration group we see a dearth of bursts in the lower left part of
the sky-distribution.
\begin{table}
\begin{tabular}{ccccc}
\hline
group       &   $N$     &   Bingham ($p$-value) & Rayleigh ($p$-value)  &   $M_N$ eigenv. $\lambda_1$ ($p$-value) \\
\hline 
short       &   $31$    &   $78.6$ ($0.644$)            & $5.6$ ($0.409$)           &   $0.387$ ($0.553$)                   \\
interm.     &   $46$    &   $123.8$ (${\bf 0.038}$)     & $10.4$ (${\bf 0.089}$)    &   $0.441$ (${\bf 0.072}$)    \\
long        &   $331$   &   $831.5$ ($0.694$)       & $23.2$ ($0.290$)      &   $0.364$ ($0.645$)                   \\
\hline
\end{tabular}
\caption{The different statistics' value for the three groups with $p$-values showing the probability of the measured value occurring by chance based on $1000$ MC simulations.}
\end{table}
         \begin{figure}[!htb]
          \includegraphics[width=.999\columnwidth]{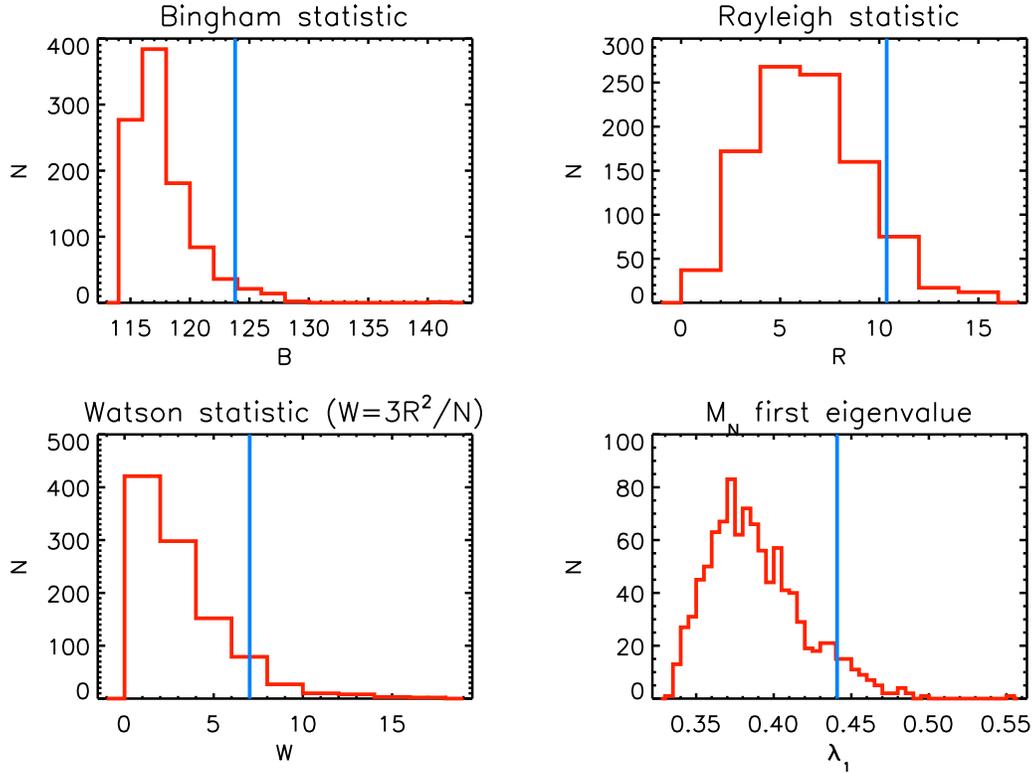}
          \caption{The distribution of 1000 MC simulated values of the four
          statistics for the intermediate-very soft group. The vertical lines mark the values of the measured data statistic. }
           \end{figure}
         \begin{figure}[htb]
          \end{figure}
\begin{theacknowledgments}
This work was supported by OTKA grant K077795, by OTKA/NKTH A08-77719 and
A08-77815 grants (Z.B.), by the GA\v{C}R grant No. P209/10/0734 (A.M.), by the
Research Program MSM0021620860 of the Ministry of Education of the Czech
Republic (A.M.) and by a Bolyai Scholarship (I.H.).
\end{theacknowledgments}
\bibliographystyle{aipproc}   
\bibliography{exposure}
\end{document}